# OPTIMAL HYBRID MULTIPLEXED AC-DC-AC POWER CONVERTERS

*Matthew Deakin [1]**

[1]*Department of Engineering, Newcastle University, Newcastle-upon-Tyne, UK*
**email: matthew.deakin@newcastle.ac.uk*

## Abstract

The flexibility of multi-terminal AC-DC-AC converters connected in distribution networks can be increased by changing the sizes of the individual AC-DC converter stages and connecting the AC side of those converters to electromechanical switches (multiplexers) to allow reconfiguration within the network. The combinations of real powers that can be transferred by such a design can be described using a capability chart. In this work, it is proposed that the area of these capability charts is a meaningful metric for describing the flexibility of such a device. These capability chart areas are calculated in closed form for a three-terminal AC-DC-AC device consisting of three AC-DC converters of arbitrary sizes, allowing the optimal AC-DC converter sizing to be determined to maximise this area. It is shown that this optimal design yields a capability chart area that is 64% larger than the equivalent area from a conventional equally-sized AC-DC-AC converter. Converters which are optimal in other senses are discussed, such as a design with 10% increased per-feeder real power transfer, albeit with an 8% area reduction. It is concluded that the capability chart area is an intuitive and informative approach for describing the increased flexibility of multiplexed AC-DC-AC converters.

## 1 Introduction

Electrical distribution systems require new, flexible capacity to enable consumers to connect low carbon technologies such as electric vehicles, heat pumps and solar photovoltaics (PV). One way of increasing this network capacity is through AC-DC-AC converters (called, amongst others, Soft Open Points, DC Links, Solid State Transformers, Smart Transformers, etc) [1]. These solutions can be installed in substations [2], in place of normally open points [1], or in parallel with switchgear [3].

It is well-known that these solutions are typically more expensive than conventional approaches of providing network capacity [2]. As a result, there has been interest in proposing Hybrid AC-DC-AC solutions that make use of low-cost electromechanical switches to increase the flexibility of these designs. For example, the Hybrid Open Point [1] installs AC-DC-AC power converters in parallel with switchgear to increase operational and planning flexibility. In the past, cost-effective hybrid approaches have allowed increased market share [4].

This work considers the novel Hybrid AC-DC-AC configuration shown in Figure 1, first introduced in [5] as the Hybrid Multi-Terminal Soft Open Point (Hybrid MTSOP). This approach uses multiplexers ('Feeder Selector Switches') to allow any converter to connect to any of the distribution feeders at a node. It is shown that this enlarges the capability chart of the device, increasing the power that can be transferred by 50% and increasing the loss reduction capabilities by 13%. The approach shows similarities with the 'MVAC switchyard' described in [6], which also uses multiplexers to reconfigure power electronics to improve network capacity. It also shows some parallels with the hybrid EV charger described in [7], which also uses a multiplexer ('relay matrix') to increase the flexibility of the outputs of a vehicle-to-grid charger, or the phase changing soft open point [13]. On the power electronics side, there is also a large literature on the design of multiport converters that make efficient use of components (e.g., reduced numbers of solid-state switches through interleaving [8] or developing systematic strategies for designing topologies with low component count [9]). However, to the best of the author's knowledge, there is no explicit quantification of the area of the increased capability charts that such a Hybrid Multiplexed AC-DC-AC system can provide. As capability charts are intuitive and well-known methods of presenting information about device flexibility, this is a significant gap.

In this paper, we address this gap by quantifying the area of the capability chart of a three-terminal Hybrid AC-DC-AC converter for any given set of three converter sizes. It is shown that it is feasible to evaluate the capability chart analytically, allowing the optimal converter design (in terms of maximum capability chart area) to be determined. A number of further cases of interest are also described to highlight the properties of the capability chart areas and more generally properties of Hybrid Multiplexed AC-DC-AC converters.

The structure of this paper is as follows. In Section 2, we define the AC-DC-AC capability chart area, to show the sense in which this metric describes operational flexibility achieved by designs using the multiplexed approach. In Section 3, we proceed to calculate these areas analytically for three combinations of systems, enabling a utility to understand the potential benefits of the approach quantitively. Finally, in Section 4 we draw salient conclusions.

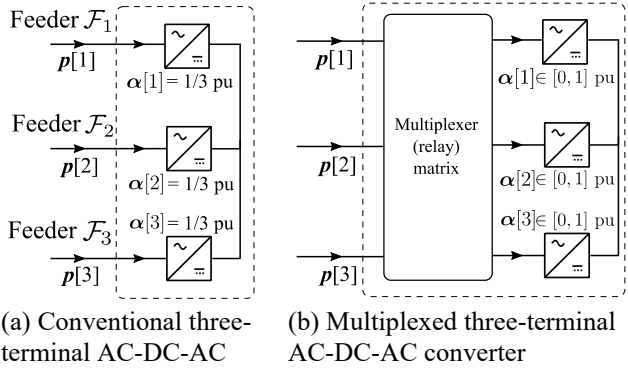

(a) Conventional three-terminal AC-DC-AC converter

(b) Multiplexed three-terminal AC-DC-AC converter

Figure 1: as compared to a conventional AC-DC-AC converter, which typically splits the power ratings equally between feeders (a), the proposed multiplexed design has asymmetrically sized converters $\alpha$ connected to feeders through a bank of multiplexers (b) to increase device flexibility.

## 2. AC-DC-AC Capability Chart Areas

The goal of this work is to evaluate the capability charts of Hybrid Multiplexed AC-DC-AC converters for a three-terminal case. In this section, the principles of the Hybrid Multiplexed AC-DC-AC converter are first outlined to give the reader a clear understanding of how multiplexing AC-DC power converters increases performance. The device capability charts are then described, and the Capability Chart Area defined mathematically to give an unambiguous description of the proposed metric that will be evaluated. Finally, necessary preliminaries are presented to give the reader a fuller understanding the results presented in Section 3.

*2.1 Principle of Operation*

The proposed Hybrid Multiplexed AC-DC-AC device was first described in [5], and so only a brief introduction to the operating principles of the device is given here. A conventional AC-DC-AC converter design would consist of three equally sized legs that are hard-wired to the feeder on which they are connected [10]. This approach has the advantage of simplicity, with the amount of capacity that can be drawn from a given feeder being fixed at 1/3 pu.

In contrast, the Hybrid Multiplexed AC-DC-AC converter has three converters, each with a different converter size, and with the AC side of the converters connected to feeders through a multiplexer, as shown in Figure 2. As in the conventional design, the total per-unit capacity of the AC-DC converters is 1 pu. However, the power that can be transferred by the device can be increased by 50% to 1/2 pu. Assuming the cost of devices is proportional to the total power capacity of the AC-DC converters, and the maximum power transferred is the limiting factor, the cost can be reduced by 33% [5].

This increase in maximum power transfer can be shown by considering a system with three AC-DC power converters, with sizes $\alpha$ = (1/2, 2/5, 1/10) pu, as shown in Figure 3. As the converter with 1/2 pu can be connected to any one of the three feeders, it can be seen that the power transfer for any feeder is increased compared to a non-configurable design with equal sizing. As well as this increase in maximum power transfer, there are also many combinations of feasible power transfers – for example, the two smaller converters can be connected in parallel to one feeder, or to different feeders (Figure 3(b)).

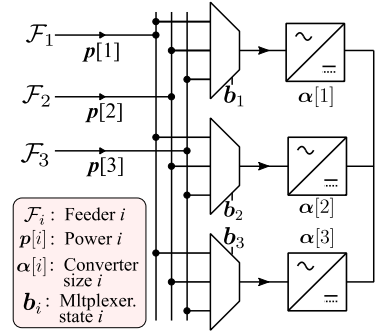

Figure 2: The three-terminal Multiplexed AC-DC-AC converter consists of three multiplexers connected to the AC side output of AC-DC converters to allow reconfiguration.

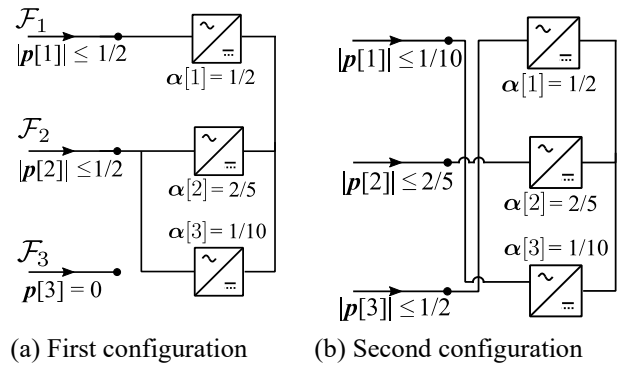

(a) First configuration   (b) Second configuration

Figure 3: Reconfiguration of power converters enables the capacity connected to each feeder to be changed depending on the needs of the network – the first configuration (a) allows 1/2 pu to be transferred only between Feeder 1 and 2, whilst the second configuration (b) allows 1/2 pu to be transferred between Feeder 3 and Feeders 1 and 2.

*2.2 Capability Charts and Capability Chart Areas*

For the purposes of this work, the combinations of power transfer that are achievable by the Multiplexed AC-DC-AC converter are considered a measure of the flexibility of the device. These combinations can be described via a Capability Chart, which can be described mathematically as follows. Let $b_i \in \{0,1\}^3$ be a vector representing the state of each multiplexer (as in Figure 2), such that

$$\sum b_i = 1,$$

and let $B \in \{0,1\}^{3\times 3}$ be a matrix concatenating these vectors, i.e., $B = [b_1, b_2, b_3]$. The capacity connected to each feeder $p^{\text{Max}} \in R^3$ is therefore

$$p^{\text{Max}} = B\alpha$$

where $\boldsymbol{\alpha} \in R^3$ is the vector of converter sizes. As this is the maximum power that a given feeder can transfer in a given configuration, the capability chart $C$ can be defined as

$$C = \{\boldsymbol{p} : \exists\, \boldsymbol{B}\, [|\boldsymbol{p}| \leq \boldsymbol{B}\boldsymbol{\alpha} \cap \sum \boldsymbol{p} = 0]\},$$

where the inequality is elementwise, and the second equality condition ensures that Kirchhoff's current law holds (for simplicity, AC-DC converters are assumed to be lossless).

Having defined the capability chart, the Capability Chart Area (CCA) is then the integral of the area of this capability chart $C$, and has units pu$^2$. This area can be denoted as

$$\text{CCA} = \int_C dA.$$

In Section 3 it will be demonstrated that this area can be evaluated directly by considering the geometry of the capability charts.

*2.2.1 Area Normalisation Constant:* The converter capability chart area is defined as an area integral over $C$. To follow the usual convention, the area should be calculated in the plane normal to the surface $C$ [11, Ch. 10.6].

It is convenient to represent the capability chart in a co-ordinate system $(\widehat{\boldsymbol{p}}_1, \widehat{\boldsymbol{p}}_2, \widehat{\boldsymbol{p}}_3)$ where each co-ordinate represents the power injected from each feeder,

$$\widehat{\boldsymbol{p}}_1 = \begin{bmatrix}1\\0\\0\end{bmatrix}, \quad \widehat{\boldsymbol{p}}_2 = \begin{bmatrix}0\\1\\0\end{bmatrix}, \quad \widehat{\boldsymbol{p}}_3 = \begin{bmatrix}0\\0\\1\end{bmatrix}.$$

However, none of these co-ordinates are orthogonal to the plane $\sum \boldsymbol{p} = 0$, as can be seen by considering the representation of this plane as

$$\boldsymbol{n}.\boldsymbol{p} = 0, \quad \boldsymbol{n} = \frac{1}{\sqrt{3}}\begin{bmatrix}1\\1\\1\end{bmatrix}.$$

To the contrary, the co-ordinates

$$\widehat{\boldsymbol{p}}_x = \frac{1}{\sqrt{2}}\begin{bmatrix}1\\0\\-1\end{bmatrix}, \quad \widehat{\boldsymbol{p}}_y = \frac{1}{\sqrt{6}}\begin{bmatrix}-1\\2\\-1\end{bmatrix}, \quad \widehat{\boldsymbol{p}}_z = \boldsymbol{n},$$

form an orthonormal co-ordinate system, with $(\widehat{\boldsymbol{p}}_x, \widehat{\boldsymbol{p}}_y)$ lying in the plane $\sum \boldsymbol{p} = \boldsymbol{0}$. Calculating the CCA in the $(\widehat{\boldsymbol{p}}_x, \widehat{\boldsymbol{p}}_y)$ frame therefore yields the conventional area.

Nevertheless, it is considered more intuitive to use $(\widehat{\boldsymbol{p}}_1, \widehat{\boldsymbol{p}}_2, \widehat{\boldsymbol{p}}_3)$ for calculating an initial area (with incorrect scaling), then correcting this in proportion to the determinant of the Jacobian of the transformation $\boldsymbol{J}$ between these spaces [11, Ch. 10.3] to find the true CCA. The determinant of this transformation is

$$|\boldsymbol{J}| = \begin{vmatrix}1/\sqrt{2} & -1/\sqrt{6}\\ 0 & \sqrt{2/3}\end{vmatrix} = \frac{1}{\sqrt{3}}.$$

Figure 4 shows geometrically the difference between representations in these co-ordinate systems for the case previously considered ($\boldsymbol{\alpha} = (1/2, 2/5, 1/10)$). In the feeder power co-ordinates $(\widehat{\boldsymbol{p}}_1, \widehat{\boldsymbol{p}}_2)$, the shape lies within a hexagon that is not regular, whilst in the orthonormal co-ordinates $(\widehat{\boldsymbol{p}}_x, \widehat{\boldsymbol{p}}_y)$ the capability chart is captured within a hexagon which is regular.

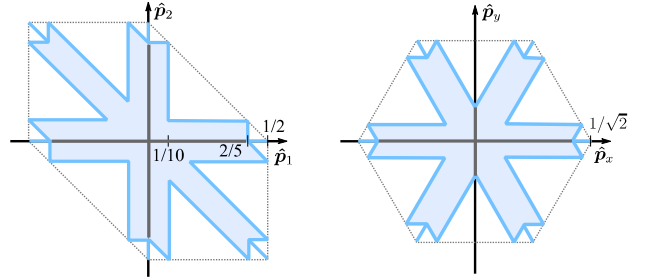

(a) Nominal co-ordinates     (b) Co-ordinates in the plane

Figure 4. The capability chart for $\boldsymbol{\alpha} = (1/2, 2/5, 1/10)$ can be plotted as (a) co-ordinates in line with each element of $\boldsymbol{p}$, or (b) co-ordinates that lie in the plane $\sum \boldsymbol{p} = 0$, with the area of a (b) factor of $\sqrt{3}$ larger than the area of (a).

*2.2.2 Ordering of Converter Sizes:* The multiplexers allow any of the three AC-DC converters to be connected to any one of the feeders. Therefore, any permutation of sizes in the converters is permissible. However, to avoid repetition (and without loss in generality), the converter sizes $\boldsymbol{\alpha}$ are ordered in decreasing size order, i.e.,

$$\boldsymbol{\alpha}[1] \geq \boldsymbol{\alpha}[2] \geq \boldsymbol{\alpha}[3] \geq 0.$$

This avoids a tiling of the plane. We also add the constraint

$$\boldsymbol{\alpha}[1] \leq \frac{1}{2},$$

because it avoids degenerate cases where a converter might have surplus capacity (e.g., if $\boldsymbol{\alpha}[1] = 0.6$, then only 0.4 pu can be passed through $\boldsymbol{\alpha}[1]$ as there is only 0.4 pu capacity split between $\boldsymbol{\alpha}[2]$ and $\boldsymbol{\alpha}[3]$). By combining these inequalities with the per-unit constraint $\sum \boldsymbol{\alpha} = 1$ we can write down that the space that needs to be considered (in the $\boldsymbol{\alpha}[1], \boldsymbol{\alpha}[2]$ plane) is a triangle defined as

$$\boldsymbol{\alpha}[1] \leq \frac{1}{2},$$
$$\boldsymbol{\alpha}[1] \geq \boldsymbol{\alpha}[2],$$
$$\boldsymbol{\alpha}[1] + 2\boldsymbol{\alpha}[2] \geq 1.$$

The final constraint can be found by noting that $\boldsymbol{\alpha}[2] \geq \boldsymbol{\alpha}[3]$, then substituting $\sum \boldsymbol{\alpha} = 1$.

If the CCA can be determined for every point in this triangle, then the size of any converter sizing $\boldsymbol{\alpha}$ can be determined (by an arbitrary permutation of indices in $\boldsymbol{\alpha}$).

# 3 Results

In the previous section, CCAs were introduced as means of capturing the total area of converter capability charts, which were proposed as a succinct method of summarising the performance of a design. In this section, we show how the CCA can be calculated from the AC-DC converter sizes $\boldsymbol{\alpha}$, allowing a complete characterization of the capability chart areas. Subsequently, we calculate the CCA of a number of cases of interest, before discussing the advantages and disadvantages of using the CCA to describe flexibility.

*3.1 Characterization of Capability Chart Areas*
The CCA for any $\boldsymbol{\alpha}$ can be found by solving

$$\beta_1 = \min\left\{\boldsymbol{\alpha}[3], \frac{\boldsymbol{\alpha}[1]}{2}\right\},$$

$$\beta_2 = \max\left\{(\boldsymbol{\alpha}[1] - \boldsymbol{\alpha}[3]), \frac{\boldsymbol{\alpha}[1]}{2}\right\},$$

$$\delta\beta = \boldsymbol{\alpha}[2] - \beta_2,$$

$$r_1 = \frac{\beta_1^2}{2},$$

$$r_2 = \beta_1(\beta_2 - \beta_1),$$

$$r_3 = \delta\beta\left(\beta_1 - \frac{\delta\beta}{2}\right),$$

$$\text{CCA} = 12\sqrt{3}(r_1 + r_2 + r_3).$$

The proof for this is shown in the Appendix, and is based on quantifying the area in the first quadrant for the three regions $r_1, r_2, r_3$, and then exploiting the geometric symmetry of the capability charts.

Figure 5 plots the CCAs for all $\boldsymbol{\alpha}$. It can be observed that the CCA varies significantly as the sizes of the converters $\boldsymbol{\alpha}$ change. For example, the maximum area of 0.945, found at $\boldsymbol{\alpha} = (0.454, 0.364, 0.182)$, has an area which is 64% greater than the conventional design, which has an area of just 0.577.

Table 1 collects and reports the areas for a number of interesting cases, each of which are plotted in Figure 6. The 'MPT optimal' case is the sizing that yields the largest area when the sizing is subject to the constraint that the converter is sized to allow the maximum power transfer (MPT) of 0.5 pu through the converter (i.e., for $\boldsymbol{\alpha}[1] = 1/2$). It can be seen that the CCA has dropped by 8% as compared to the design with the largest CCA, but this design comes with the benefit that the maximum power that can be transferred has been increased by more than 10% from 0.454 to 0.5 pu. Depending on the purposes of the converter in the system, one or the other of these designs may be preferable. For example, if the converter is regularly used at its maximum rating, potentially the 'MPT optimal' design may be preferable, where the 'Optimal' design may be a better design when the power transferred is more variable.

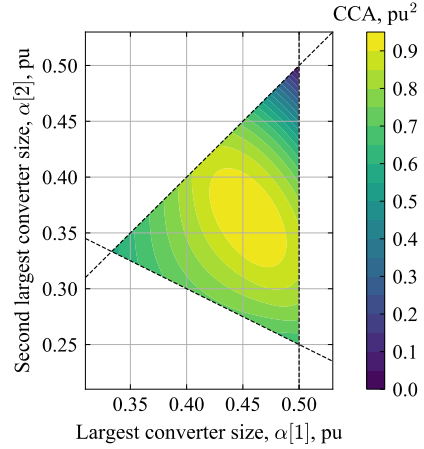

Figure 5. The Converter Capability Area as a function of converter sizes $\boldsymbol{\alpha}[1], \boldsymbol{\alpha}[2]$ (with $\boldsymbol{\alpha}[3]$ implied as $\sum \boldsymbol{\alpha} = 1$ ).

Table 1 Converter capability chart areas for several designs

| Design | Converter sizes $\boldsymbol{\alpha}$, pu | CCA, pu$^2$ |
|---|---|---|
| Conventional | (1/3, 1/3, 1/3) | 0.577 |
| Optimal | (0.454, 0.364, 0.182) | 0.945 |
| MPT optimal | (1/2, 1/3, 1/6) | 0.866 |
| Convex optimal | (0.4, 0.4, 0.2) | 0.831 |
| Two converters [6] | (1/2, 1/2, 0 ) | 0 |
| Perfect converter | na | 1.299 |

Table 1 also reports the CCAs for the 'Convex optimal' converter. This is the converter which has the largest CCA which is convex (along the line $\boldsymbol{\alpha}[1] = \boldsymbol{\alpha}[2]$). This is an interesting converter sizing because it is a linear enlargement of the Conventional design, as can be seen in Figure 6. This means that a network operator could model the capability chart for design identically to the conventional design, but in the network itself the device would have to make use of the switches to achieve the power flows requested from the network operator.

Finally, the table reports the CCA for the 'Two converter' and 'Perfect converter' designs. The former consists of just two converters with 1/2 pu capacity on each converter (still with the AC-DC converters having multiplexers on the output). Trivially, this has zero area. Nevertheless, this could be a useful design, as it allows larger amounts of power transfer than, say, the Conventional design, and would have fewer moving parts. The 'Perfect converter' is not a physical converter sizing, but is rather the boundary of possible power transfers from the device (i.e., the locus traced if $\boldsymbol{p}[i] = 0.5$ pu for each feeder in turn), and is included for comparative purposes (this is plotted as the dashed line in Figure 4 and Figure 6).

*3.2 Discussion*
In this work, we have considered the transfer of real power through the device. However, if a Hybrid AC-DC-AC converter is installed in a real network, it is possible that the device might also be used to provide reactive power to reduce losses, provide voltage support or improve power factor. In that case, for a three-feeder system, the dimension of the capability chart would increase by 3 (the reactive power injected into each feeder is independent of other feeders, unlike the real power, which is constrained by $\sum p = 0$). The much higher dimensionality would make the evaluation of the volume of such a 'converter capability hypervolume' much more challenging, and so is beyond the scope of this work. It is noted, however, that techniques such as Monte Carlo Integration can be used to evaluate these multidimensional integrals numerically [12, Ch. 4.8].

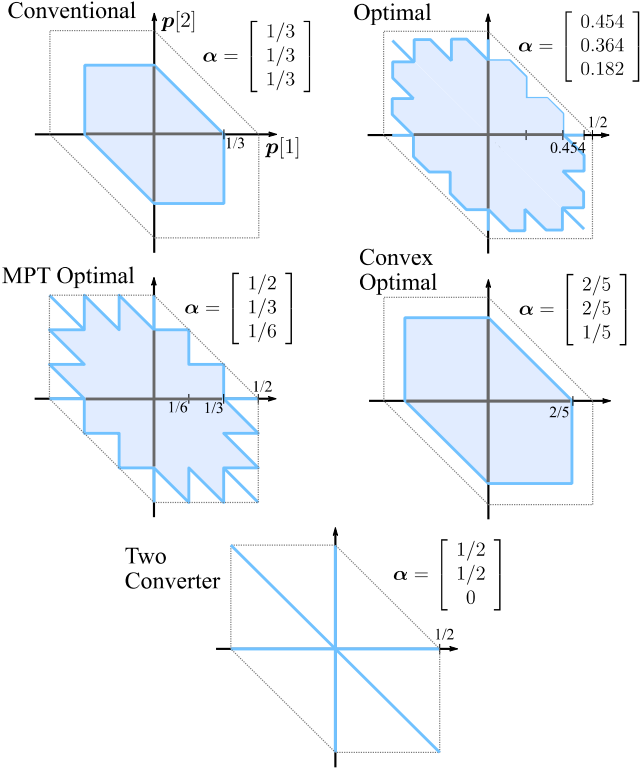

Figure 6. The capability charts of the five AC-DC-AC designs considered in Table 1.

The CCA does not consider real-world issues that would affect the practicality of such a device, such as potential reduction in device reliability and increased space requirements due to the electromechanical multiplexer switches. Design choices could be made to reduce numbers of these switches to reduce the number of moving parts and volume required to address these issues, whilst still achieving good CCA.

Finally, the strength of the link between capability chart areas and actual use in distribution networks needs further investigation. In the case study considered in [5], it was shown that the Two converter case, sized as $\alpha = (0.5, 0.5, 0)$, has better performance than the Conventional sizing approach; however, the CCA is trivially zero for the former design. Clearly, the CCA is not the only determinant of performance. It is therefore suggested that further work could consider alternative characterisations of the flexibility of the Hybrid AC-DC-AC converters.

## 4 Conclusion

Power converters will be an integral part of future power systems, and AC-DC-AC converters have been proposed to provide a wide range of network services in a variety of contexts. The utilization of these power converters can be increased by hybridising them, using multiplexers on the DC-AC output to allow these converters to be flexibly switched between distribution feeders. The Capability Chart Area was introduced as a metric for characterising the flexibility of different hybrid AC-DC-AC designs, according to the area of the individual AC-DC converters.

The area of this chart has been fully characterised for a three-terminal hybrid AC-DC-AC converter, showing an increase in this area of 64% for the optimal case as compared to the conventional, equally sized converter case. There are, however, trade-offs, with some converter designs allowing increased maximum power transfer between feeders for a small reduction in this area. It is concluded that the simplicity and conciseness of the capability chart area lends it to be an informative and useful metric for evaluating and comparing the flexibility of both multiplexed and hard-wired, non-configurable AC-DC-AC converters designs.

## 5 Appendix: Determining the CCA

The approach used to determine the CCA is based on three steps. Firstly, the space is split into twelve, according to the upper and lower halves of each 'arm' of the hexagonal capability charts (in $(\hat{p}_1, \hat{p}_2)$ co-ordinates, as described in Section 2.2.1). Secondly, the area of one of these halves of an 'arm' is calculated (the arm in the lower half of the first quadrant is considered). Finally, this individual area multiplied by the number of half-arms (12) and the linear scaling coefficient ($\sqrt{3}$, as described in Section 2.2.1) determines the total CCA.

The only non-trivial step is the second step, determining the area of the capability chart for one arm half. To do so, the area of the capability chart for one half-arm is again split into three regions, as shown in Figure 7. In this half quadrant, it has been assumed that the smaller converter $\alpha[3]$ is connected to Feeder 2, the largest converter $\alpha[1]$ connected to Feeder 3, and the final converter $\alpha[2]$ is connected to Feeder 1. These reason these need to be connected in this order is as follows.

- Firstly, note that for a point to lie above the line $p[2] = 0$ (so that it contributes to the CCA), two converters cannot be connected in parallel to a single feeder.

- The smallest converter $\alpha[3]$ is connected to Feeder 2 as this has the smallest or equal smallest power - $p[1] \geq p[2]$ by definition; $|p[3]| \geq 2p[2]$, by considering the previous inequality and $\sum p = 0$.

- Finally, there are then two options for the final converter connection. If the largest converter $\alpha[1]$ is connected to Feeder 1 then $p[1] \leq \alpha[1]$ and $|p[1] + p[2]| \leq \alpha[2]$; otherwise, $p[1] \leq \alpha[2]$ and $|p[1] + p[2]| \leq \alpha[1]$. The latter is a larger area.

The three regions are then defined as shown in Figure 7, consisting of a triangular area $r_1$, a rectangular area $r_2$, and then a final trapezoidal area $r_3$. By inspection, the value of $\beta_1$ is the smaller of $\alpha[3]$ and $\alpha[1]/2$. Similarly, the value of $\beta_2$ is the larger of $\alpha[1]/2$ and the intersection between $p[2] = \alpha[1]/2$ and $p[1] + p[2] = \alpha[1]$ (i.e., $\alpha[1] - \alpha[3]$). This concludes the proof.

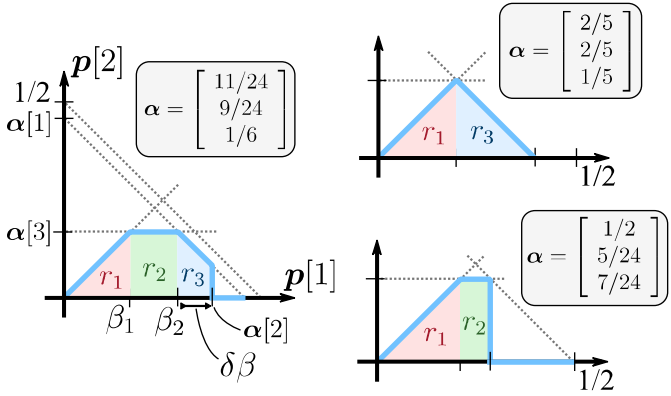

Figure 7. The three regions used to calculate the CCA described in Section 3.1, as defined by $\beta_1, \beta_2$ and $\delta\beta$ (left). Regions $r_2, r_3$ maybe have zero area, depending on the AC-DC converter sizing $\alpha$.

## 5 Acknowledgements

The author wishes to thank Dr Arman Alahyari for providing valuable comments on this manuscript. This work was supported the Centre for Postdoctoral Development in Infrastructure, Cities and Energy (C-DICE) programme led by Loughborough University in partnership with Cranfield University and the University of Birmingham. C-DICE is funded by the Research England Development Fund. This work was also support by the Multi-energy Control of Cyber-Physical Urban Energy Systems (MC2) project (EP/T021969/1).